\documentclass[12pt,fleqn]{article}
\usepackage{graphicx}
\begin{document}

\title{\Large Thermodynamic curvature measures interactions}

\author{
   George Ruppeiner\footnote{ruppeiner@ncf.edu}\\
   Division of Natural Sciences\\
   New College of Florida\\
   5800 Bay Shore Road\\
   Sarasota, Florida 34243-2109 }

\maketitle

\renewcommand{\baselinestretch}{1.0}

\begin{abstract}

Thermodynamic fluctuation theory originated with Einstein who inverted the relation $S=k_B\ln\Omega$ to express the number of states in terms of entropy: $\Omega= \exp(S/k_B)$. The theory's Gaussian approximation is discussed in most statistical mechanics texts. I review work showing how to go beyond the Gaussian approximation by adding covariance, conservation, and consistency. This generalization leads to a fundamentally new object: the thermodynamic Riemannian curvature scalar $R$, a thermodynamic invariant. I argue that $|R|$ is related to the correlation length and suggest that the sign of $R$ corresponds to whether the interparticle interactions are effectively attractive or repulsive.

\end{abstract}

\section{INTRODUCTION}

\renewcommand{\baselinestretch}{1.0}

Thermodynamic properties follow from particle properties via statistical mechanics. But thermodynamics is more than an application of statistical mechanics. Thermodynamics can sometimes yield microscopic information by means of the fluctuation theory as originated by Einstein in 1904 \cite{Ein}.

What does thermodynamic fluctuation theory tell us? In its well-known Gaussian approximation it yields the probability of finding a system in some thermodynamic state. Most texts do not go further than this Gaussian approximation. However, here I argue that going further using ideas from Riemannian geometry adds a fundamentally new element to the picture: the thermodynamic Riemannian curvature $R$, which yields information about interparticle interactions \cite{Rupp95}.

I start by examining thermodynamic fluctuations with a single variable, and emphasize the problems that result on attempting to go beyond the Gaussian approximation. A corrected theory is given, and when a second independent fluctuating variable is added to this theory the meaning of $R$ emerges. $|R|$ gives the correlation length and the sign of $R$ corresponds to whether the interparticle interactions are effectively attractive or repulsive.

\section{ONE FLUCTUATING VARIABLE}

Consider a thermodynamic system consisting of a set of microscopic constituents contained in a box of volume $V$ with an internal energy $U$. Imagine that we may specify the thermodynamics completely by writing the system's entropy $S$ as a function of $U$ and $V$,
\begin{equation}
S=S(U,V), \label{-10}
\end{equation}
the fundamental equation \cite{Call}. Examples for which $S(U,V)$ is known exactly are a gas of electromagnetic radiation (photons) and a paramagnet of noninteracting spins. Although the fundamental equation is rigorous only in the thermodynamic limit of infinite volume, thermodynamic applications use Eq. (\ref{-10}) for systems of any volume. Because $S$, $U$, and $V$ are all additive, the fundamental equation may be written as
\begin{equation} S=Vs(u), \label{-20}
\end{equation}
where $\{s,u\}\equiv\{S/V,U/V\}$ are quantities per volume \cite{Call}.

Let this thermodynamics constitute a large, closed universe $A_0$ with constant internal energy $U_0$ and constant volume $V_0\rightarrow\infty$. Imagine $A_0$ to be separated into two parts joined by an immovable, open partition enclosing a finite system $A$, with fluctuating internal energy $U$ and constant volume $V$. The rest of the universe is the reservoir $A_r$, with fluctuating internal energy $U_r$ and constant volume $V_r\rightarrow\infty$ as illustrated in Fig. 1. Quantities with subscripts $0$ and $r$ refer to properties of $A_0$ and $A_r$, respectively. From Eq. (\ref{-20}) we know that a single entropy per volume function $s=s(u)$ is common to all three systems $A_0$, $A_r$, and $A$.

$A$ and $A_r$ exchange energy through fluctuations, which are described by thermodynamic fluctuation theory. The microcanonical ensemble asserts that all accessible microstates of $A_0$ occur with equal probability \cite{Landau}. Therefore, the probability of finding the internal energy per volume of $A$ between $u$ and $u+du$ is proportional to the number of microstates of $A_0$ corresponding to this range:
\begin{equation} P(u,V)du=C \Omega_0(u,V)du, \label{-25}
\end{equation}
where $\Omega_0(u,V)$ is the density of states, and $C$ is a normalization factor. Boltzmann's expression for the entropy $S=k_B \ln \Omega$, where $k_B$ is Boltzmann's constant, yields Einstein's famous relation \cite{Landau}:
\begin{equation} P(u,V)du=C\exp\left[\frac{S_0(u,V)}{k_B}\right]du, \label{-30}
\end{equation}
where $S_0(u,V)$ is the entropy of $A_0$ when $A$ has internal energy per volume $u$. Equation (\ref{-30}) is the basis of thermodynamic fluctuation theory.

Because entropy is additive, we have
\begin{equation}
S_0(u,V) = V s(u) + V_r \,s(u_r). \label{-40}
\end{equation}
In statistical mechanics we assume that there is some unique division of energy between $A$ and $A_r$ that maximizes $S_0(u,V)$, corresponding to equilibrium. Let this maximum correspond to $u=u^*$ and $u_r=u_r^*$, and define the differences $\Delta u=u-u^*$ and $\Delta u_r=u_r-u_r^*$. We expand both entropy densities in Eq. (\ref{-40}) about the maximum and obtain
\begin{eqnarray}
S_0(u,V) &=& \tilde{S}_0 + V s'(u^*) \Delta u + V_r s'(u_r^*)\Delta u_r + V\frac{1}{2!} s''(u^*)(\Delta u)^2 \nonumber \\ &&{} + V_r\frac{1}{2!} s''(u_r^*)(\Delta u_r)^2 + V\frac{1}{3!} s'''(u^*)(\Delta u)^3 + \cdots, \label{-50}
\end{eqnarray}
where the prime indicates differentiation with respect to $u$, and $ \tilde{S}_0$ is the entropy of $A_0$ in equilibrium.

Expansion about a maximum means that the first-order terms in Eq. (\ref{-50}) must add to zero. By conservation of energy,
\begin{equation}
V\Delta u=-V_r\Delta u_r\label{-55},
\end{equation}
and the cancellation of the first-order terms requires that $s'(u^*) = s'(u_r^*)$. This cancellation forces $u^*= u_r^*$ because the inverse temperature $s'(u)$ is assumed to decrease monotonically with increasing $u$. Additivity of energy further requires that $V u^*+V_r u_r^*=V_0 u_0$, leading to $u^*= u_r^*=u_0$. The differences now become $\{\Delta u,\Delta u_r\}=\{u-u_0,u_r-u_0\}$. Energy conservation expressed in Eq. (\ref{-55}) also means that terms beyond first order in Eq. (\ref{-50}) from expanding $s(u_r)$ will be negligible compared with the corresponding terms from expanding $s(u)$. Hence,
\begin{equation} S_0(u,V) = \tilde{S}_0 + V\frac{1}{2!} s''(u_0)(\Delta u)^2 +V\frac{1}{3!} s'''(u_0)(\Delta u)^3 + \cdots. \label{-60}
\end{equation}

Truncating this series at second order leads to the Gaussian approximation in thermodynamic fluctuation theory \cite{Landau,Pathria}:
\begin{equation} P_G(u,V)du = \sqrt{\frac{V}{2\pi}} \exp\left[-\frac{V}{2}g(u_0) (\Delta u)^2\right]\sqrt{g(u_0)}\,du, \label{-70}
\end{equation}
where
\begin{equation} g(u_0)\equiv \frac{-s''(u_0)}{k_B} \label{-75}
\end{equation}
is always positive because it is proportional to the heat capacity. The normalization factor in Eq. (\ref{-70}) results from Gaussian integration, which also yields the averages
\begin{equation} \left<\Delta u\right> = \int \Delta u\,P_G(u,V) du = 0,\label{-80}\end{equation}

\noindent and

\begin{equation} \left<\left(\Delta u\right)^2\right> = \int \left(\Delta u\right)^2 P_G(u,V) du=\frac{1}{Vg(u_0)}. \label{-90}\end{equation}

Let's move on to some points not found in books on statistical mechanics. First, although the conservation rule $\left< \Delta u\right>=0$ results almost trivially from the Gaussian integration Eq. (\ref{-80}), it is easy to construct an argument that it must hold for any physically correct probability density $P(u,V)$ for all $V$ \cite{footnote1}. However, beyond the Gaussian approximation, thermodynamic fluctuation theory does not obey the conservation rule. For decreasing $V$, fluctuations in $\Delta u$ increase by Eq. (\ref{-90}), and discarding the third-order term in the entropy expansion in Eq. (\ref{-60}) becomes increasingly difficult to justify. If in Eq. (\ref{-80}) we were to calculate the integral with $P(u,V)$ containing a third-order term, we would find $\left<\Delta u\right>\not= 0$, because the probability density is no longer symmetric about $u=u_0$. $\left<\Delta u\right>\not= 0$, violates the required conservation rule.

Another problem associated with going beyond the Gaussian approximation becomes evident on changing the thermodynamic coordinate $u$ to a general coordinate $x(u)$. In the microstate counting argument which led to Eq. (\ref{-30}), no special properties of $u$ were used (conservation of energy comes only later in the argument). We could as well have used any other thermodynamic function $x=x(u)$ and written
\begin{equation}
P(x,V)dx=\tilde{C}\exp\left[\frac{S_0(x,V)}{k_B}\right]dx, \label{-100}
\end{equation}
where $\tilde{C}$ is a normalization factor. But, innocuous as it seems, Eq. (\ref{-100}) is inconsistent with Eq. (\ref{-30}). If we transform coordinate $u$ to $x(u)$ on the left-hand side of Eq. (\ref{-30}), we obtain
\begin{equation} P(u,V)du = \left[P(u,V)\left(\frac{du}{dx}\right)\right]dx \equiv P(x,V)dx. \label{-102}
\end{equation}
On the right-hand side of Eq. (\ref{-30}) the function $S_0(u,V)=S_0(x,V)$ remains the same under this transformation because entropy is a function of state, but $du=(du/dx)dx$ picks up the Jacobian $J=(du/dx)$. Because $J$ varies with the thermodynamic state, it cannot simply be absorbed into the normalization factor, and we end up with
\begin{equation} P(x,V)dx = C \left(\frac{du}{dx}\right) \exp\left[\frac{S_0(x,V)}{k_B}\right]dx, \label{-103}
\end{equation}
which is inconsistent with Eq. (\ref{-100}).

We call an inconsistency resulting from a change in coordinates a lack of covariance \cite{Rupp95}. It means that either Eq. (\ref{-30}) or Eq. (\ref{-100}) or both are incorrect. It is significant that this problem does not arise in the Gaussian approximation. If we change the coordinate in Eq. (\ref{-70}), we obtain the transformed Gaussian approximation for the $x$ coordinate
\begin{equation}
P_G(x,V)dx = \sqrt{\frac{V}{2\pi}} \exp\left[-\frac{V}{2}\,g(x_0)(\Delta x)^2\right]\sqrt{g(x_0)}\,dx, \label{-105}
\end{equation}
where the transformed function $g(x_0)$ is
\begin{equation} g(x_0)=g(u_0)\left(\frac{du}{dx}\right)^2, \label{-106}
\end{equation}
with $\Delta x=x-x_0$ and $x_0=x(u_0)$. The Gaussian approximation is clearly covariant as long as we use the transformation rule Eq. (\ref{-106}). Of course, we require small fluctuations for its validity.

Coming back to the $u$ coordinate, observe that if $g(u)$ were a constant function, then all derivatives of $s(u)$ from third order and up would be zero, and the Gaussian approximation, which is covariant and obeys the conservation rule, is exact for all $V$. The function $g(u)$ is given by the thermodynamics of the system of interest and thus we have no control over whether or not it is constant. But could we transform to another thermodynamic coordinate $x=x(u)$ where the function $g$ in the Gaussian approximation is constant, and higher order terms do not appear? Addressing this question is a major theme of this paper. But the answer is not to be found in Eq. (\ref{-30}), which is fundamentally unsound beyond the Gaussian approximation. We need a new approach.

We start by adding a second, larger partition concentric with the first. This partition encloses a system $A_1$ of which $A$ is part as shown in Fig. 2. The infinite reservoir $A_r$ is now everything in $A_0$ outside of $A_1$. Because $A_1$ is larger than $A$, fluctuations in $\Delta u_1$ will be less than those of $\Delta u$, by Eq. (\ref{-90}), and the Gaussian approximation should work better for it. We now calculate the fluctuation probability of $u$ about $u_1$ by imagining $A_1$ to be in a state with energy per volume $u_1$. Consider the reservoir surrounding $A$ to be system $A_1-A$, rather than the infinite $A_r$. Expand the entropy of $A_1$ about its maximum as in Eq. (\ref{-50}). Again, the first-order terms drop out, but, in contrast to the case with an infinite reservoir, we must keep both quadratic terms. I omit the details here (see Ref. \cite{Rupp95}). The key is that the resulting fluctuations in $\Delta u$ about $u_1$ are smaller than the corresponding ones about $u_0$, with the state of $A_1$ unspecified, and the Gaussian approximation should thus be improved.

Two probability distributions ($A$ fluctuating about $A_1$ and $A_1$ fluctuating about $A_0$) are usually more than we ultimately need. Multiplying the two probability distributions and integrating out the intermediate state $u_1$ of $A_1$, gets us what we want: the probability of $A$ fluctuating about $A_0$. The result should be better than that resulting from Eq. (\ref{-70}) alone, because the pair of Gaussian approximations are each improved, and because in the integral with respect to $u_1$ we can evaluate $g(u)$ at the variable $u_1$ rather than always at the fixed $u_0$.

We can improve on this two probability structure by going to a continuum limit with more concentric partitions, filling $A_0$ densely with them from very large down to $A$. Each system fluctuates about the state of the one immediately larger than it according to a Gaussian approximation. To obtain the probability of finding $A$ in some state, given the state of $A_0$, we integrate out all the intermediate states. Such a multiple integration is called a path integral. To make it covariant and have it satisfy the conservation rule for all $V$ requires great care. I skip the difficult details here \cite{Grabert}, and approach this problem instead from another direction.

Let's work on finding a partial differential equation for $P(u,V)$. In many cases in physics, fundamental principles obey partial differential equations, and there is no reason why this would not be the case here. Finding an acceptable partial differential equation much limits the possible ways the modified thermodynamic fluctuation theory could be expressed. To this end, note first that the form of the Gaussian approximation Eq. (\ref{-70}) looks like the solution of a diffusion equation, with the role of ``time'' played by the inverse volume
\begin{equation}
t\equiv\frac{1}{V}, \label{-108}
\end{equation}
and with a Dirac delta function ``initial condition'' ($t=0$) peaked at $\Delta u=0$. I now explore this connection in a general coordinate $x=x(u)$, and will return to the coordinate $u$ later.

The most general form of the diffusion equation which preserves normalization \cite{footnote2} may be written as \cite{Rupp95,Graham}
\begin{equation} \frac{\partial P}{\partial t}=-\frac{\partial}{\partial x}[K(x)P] + \frac{1}{2}\frac{\partial^2}{\partial x^2}\left[g^{-1}(x)P\right], \label{-110}
\end{equation}
where the drift term $K(x)$ and $g(x)$ are both functions of $x$ to be determined. We understand that nothing in our problem is diffusing in time; $t$ is simply a measure of volume, and the ``diffusion'' reflects the fact that the thermodynamic state of $A$ becomes increasingly uncertain as $V$ decreases, and the probability distribution for $u$ gets broader, by Eq. (\ref{-90}). If we can determine $K(x)$ and $g(x)$, then they, along with the initial condition and appropriate boundary conditions, complete the covariant theory \cite{Rupp95}.

For constant functions $K(x)=K$ and $g(x)=g$ it is straightforward to verify that the exact normalized solution to Eq. (\ref{-110}) is the Gaussian expression
\begin{equation} P_G(x,t)=\sqrt{\frac{g}{2\pi t}} \exp\left[-\frac{1}{2t}\,g(\Delta x-K t)^2\right], \label{-120}
\end{equation}
where $\Delta x\equiv x-x_0$, and $x_0$ is a constant. This solution has the Dirac delta function $\delta(x-x_0)$ start at $t=0$. We can also readily verify the Gaussian integrals

\begin{equation} \left<\Delta x\right> = \int \Delta x\,P_G(x,t) dx = K t,\label{-130}\end{equation}

\noindent and

\begin{equation} \left<\left(\Delta x\right)^2\right> = \int \left(\Delta x\right)^2 P_G(x,t) dx=\frac{t(1+g K^2 t)}{g}, \label{-140}\end{equation}

\noindent both valid for all $t$.

For small $t$ the Gaussian expression in Eq. (\ref{-120}) peaks sharply about the starting value $x_0$, and $P_G(x,t)$ does not sample the functions $K(x)$ and $g(x)$ over any substantial range that deviates from $x_0$. Taking $K(x)$ and $g(x)$ to be constant (and evaluated at $x_0$) in Eq. (\ref{-110}) is thus justified for small $t$. Deviations of the true $P(x,t)$ from the Gaussian expression is related to $P(x,t)$ being spread over a large enough range that we can no longer consider $K(x)$ and $g(x)$ to be constant in solving Eq. (\ref{-110}).

The Gaussian integrals above connect both $K$ and $g$ to the thermodynamics of the system under consideration. Take $t$ to be small, and switch back to the $u$ coordinate. Compare Eqs. (\ref{-120}), (\ref{-130}), and (\ref{-140}) with the corresponding ones from the thermodynamic Gaussian expression Eqs. (\ref{-70}), (\ref{-80}), and (\ref{-90}). Observe that we must have $K=0$ and $t=1/V$. Also observe that $g$ is common to the two cases, is given by Eq. (\ref{-75}), and determines the width of the fluctuations. At first sight it might appear that the function $K(x)$ in Eq. (\ref{-110}) is irrelevant because in $u$ coordinates $K(u)=0$. But nonzero $K(x)$ turns out to be necessary to transform coordinates correctly.

To make further progress with the theory, we must learn how to transform Eq. (\ref{-110}) to another coordinate $\tilde{x}=\tilde{x}(x)$. First we must ask, is it even possible to transform $P(x,t)$, $K(x)$, and $g(x)$ such that the form of the partial differential equation is unchanged:
\begin{equation} \frac{\partial \tilde{P}}{\partial t}=-\frac{\partial}{\partial \tilde{x}}\left[\tilde{K}(\tilde{x})\tilde{P}\right] + \frac{1}{2}\frac{\partial^2}{\partial \tilde{x}^2}\left[\tilde{g}^{-1}(\tilde{x})\tilde{P}\right]? \label{-150}
\end{equation}
That the partial differential equation keeps its form is required for covariance. Because probability is a scalar quantity independent of the choice of coordinate, we require $\tilde{P} d\tilde{x}=P dx$, and thus
\begin{equation}
\tilde{P}=P\left(\frac{dx}{d\tilde{x}}\right). \label{-160}
\end{equation}
We substitute this expression into Eq. (\ref{-150}), replace $\partial/\partial\tilde{x}$ by $(dx/d\tilde{x})\partial/\partial x$, divide by $dx/d\tilde{x}$, and equate coefficients of the corresponding derivatives of $P$ with those in Eq. (\ref{-110}). The result is equality if and only if the functions $K$ and $g$ transform as
\begin{equation}
\tilde{g} =g\left(\frac{dx}{d\tilde{x}}\right)^2, \label{-170}
\end{equation}
\noindent and
\begin{equation}
\tilde{K}=K \frac{d\tilde{x}}{dx}+\frac{1}{2}\,g^{-1}\frac{d^2 \tilde{x}}{dx^2}. \label{-180}
\end{equation}
Hence, the partial differential equation Eq. (\ref{-110}) is covariant if we use the appropriate transformation rules. I add that it was not a priori clear that we could find acceptable transformation rules allowing covariance. Our success further attests to the validity of Eq. (\ref{-110}). The transformation rules and the connection to thermodynamics in the $u$ coordinate determine $K(x)$ and $g(x)$ uniquely for any coordinate $x=x(u)$. We have thus solved the problem of fluctuation theory with one variable.

Is it possible to transform to a coordinate $\tilde{x}=\tilde{x}(u)$ with transformed $\tilde{K}(\tilde{x})$ and $\tilde{g}(\tilde{x})$ constant everywhere? If so, the Gaussian expression Eq. (\ref{-120}) would apply for all $t$. First observe that finding $\tilde{x}=\tilde{x}(u)$ with constant $\tilde{g}$ is always possible. By Eq. (\ref{-170}) we see that
\begin{equation}\tilde{x}=\sqrt{\tilde{g}^{-1}} \!\int\!   \sqrt{g(u)}\,du,     \label{-190}
\end{equation}
with any constant $\tilde{g}$, does the job. However, knowing $\tilde{x}(u)$ leaves us with no more degrees of freedom, and we generally cannot also solve Eq. (\ref{-180}) for constant $\tilde{K}$. Our attempt at a perfect coordinate is only half successful. But notice that for small $t$ it is $\tilde{g}$ which determines the width of the fluctuations; $\tilde{K}$ is needed only as a correction term to keep the average energy fixed as $t$ becomes bigger (and is not even relevant for small $t$). Thus, although we have solved only half the problem, we have solved the better half. Regardless, a necessary condition for the Gaussian expression to hold for all $t$, constant $\tilde{g}$, may always be satisfied with one variable. 

I mention in passing another theme in addition to covariance and conservation of energy. There is more than one way to obtain $P(x,t)$ from Eq. (\ref{-110}), and the various ways must yield the same results -- the requirement of consistency. The first (and best) way is to solve for $P(x,t)$ in a single step, starting from a delta function initial condition at $t=0$. Also possible is a two step solution method, as in Fig. 2, where we first solve for $P(x_1,t_1)$ from $t=0$ to some $t_1<t$ for the probability of finding $x_1$. Then we solve to $t$ for the conditional probability $P(x, t|x_1,t_1)$, the probability that we get $x$ at $t$ given that we know it is $x_1$ at $t_1$. Integrating out the intermediate variable yields $P(x,t)=\!\int P (x, t|x_1,t_1)P(x_1,t_1)dx_1$. To be consistent, $P(x,t)$ must agree with that from the single step method.

I point out that solutions to Eq. (\ref{-110}) satisfy covariance, conservation, and consistency \cite{Rupp95}, in contrast to $P(x,t)$ calculated from classical thermodynamic fluctuation theory, Eq. (\ref{-30}), which satisfies none of these conditions beyond the Gaussian approximation. I claim that the theory based on Eq. (\ref{-110}) is the correct way to extend thermodynamic fluctuation theory beyond the Gaussian approximation \cite{Rupp83a,Rupp83b,Diosi}. Ruppeiner \cite{Rupp95} has given an explicit calculation for the simple paramagnet, including the issue of boundary conditions (not considered here). The covariant theory shows considerable improvement over the classical one.

The improvements resulting from the covariant theory with just one variable affect little in physics, because problems characterized by fundamental equations of the form $S=S(U,V)$ are generally well understood. But, the story changes if we add another variable. A fundamentally new object appears, the thermodynamic curvature.

\section{TWO FLUCTUATING VARIABLES}

Consider a pure fluid system (gas or liquid) of $N$ identical particles in a volume $V$ which may be open or closed. The fundamental equation is \cite{Call}
\begin{equation} S=S(U,N,V).\label{50}
\end{equation}
Because $S$, $U$, $N$, and $V$ are all additive, we may write instead
\begin{equation}
S=V s(u,\rho),\label{60}
\end{equation}
where $\{s,u,\rho\}\equiv\{S/V,U/V,N/V\}$ are quantities per volume.

We again partition a closed, infinite universe $A_0$ into two parts, a finite $A$ and an infinite reservoir $A_r$, by a fixed, imaginary partition as shown in Fig. 1. Both $U$ and $N$, but not $V$, fluctuate \cite{Landau}. Define $\{a^1,a^2\} \equiv\{u,\rho\}$, and
\begin{equation} F_\alpha \equiv \frac{\partial s}{\partial a^\alpha},\label{70}
\end{equation}
where $\alpha=1,2$. Basic thermodynamics gives $\{F_1,F_2\}=\{1/T,-\mu/T\}$, where $T$ is the temperature, and $\mu$ is the chemical potential. The properties of $A_0$ and $A_r$ are denoted by subscripts $0$ and $r$, respectively.

The probability of finding the state of $A$ in the range $(u,\rho)$ to $(u+du,\rho+d\rho)$ is
\begin{equation} Pdu d\rho=C \exp\left(\frac{S_0}{k_B}\right)du d\rho,\label{80}
\end{equation}
where $C$ is a normalization factor, and $S_0$ may be written as
\begin{equation} S_0 = V s(u,\rho) + V_r s(u_r,\rho_r).\label{90}
\end{equation}
If we apply the same argument as used in Sec. II, there will be a maximum for $S_0$ in the state where $a^{\alpha}= a^{\alpha}_r=a^{\alpha}_0$. I expand the entropies in Eq. (\ref{90}) about this maximum in powers of the differences $\Delta a^\alpha=a^{\alpha}-a^{\alpha}_0$ and $\Delta a^{\alpha}_r=a^{\alpha}_r-a^{\alpha}_0$ and obtain\begin{eqnarray} \Delta S_0 &=& V F_\mu \Delta a^\mu + V_r F_{r\mu} \Delta a_r^\mu + V\frac{1}{2!}\,\frac{\partial F_{\mu}}{\partial a^\nu}\Delta a^\mu\Delta a^\nu \nonumber \\ &&{}+ V_r\,\frac{1}{2!}\,\frac{\partial F_{r\mu}}{\partial a_r^\nu}\Delta a_r^\mu\Delta a_r^\nu + \cdots, \label{100}
\end{eqnarray}
where $\Delta S_0$ is the difference between $S_0$ and its maximum value, and summation over repeated indices is assumed.

Conservation of both energy and particles requires $V \Delta a^{\alpha}=-V_r\Delta a_r^{\alpha}$. A necessary condition for maximum entropy is thus $F_\alpha = F_{r\alpha}$. Also, the second quadratic term in Eq. (\ref{100}) is negligible compared with the first, and Eq. (\ref{100}) becomes
\begin{equation} \frac{\Delta S_0}{k_B} =- \frac{V}{2}g_{\mu\nu}\Delta a^\mu \Delta a^\nu,\label{140}
\end{equation}
where the symmetric intensive matrix
\begin{equation} g_{\alpha\beta} \equiv -\frac{1}{k_B}\frac{\partial^2 s}{\partial a^\alpha\partial a^\beta} \label{150}
\end{equation}
must be positive-definite because $\Delta S_0\le0$. (This property holds only for short-range interactions typical of most fluid and magnetic systems. If, for example, gravity is present, it might be difficult to find a local entropy maximum.)

If we substitute Eq. (\ref{140}) into Eq. (\ref{80}) and calculate the normalization factor, we obtain the Gaussian approximation \cite{Landau}
\begin{equation} P_G \, da^1da^2 = \Big(\frac{V}{2\pi} \Big) \exp \Big(-\frac{V}{2}g_{\mu\nu}\Delta a^\mu \Delta a^\nu\Big)\sqrt{g}\,da^1da^2, \label{154}
\end{equation}
where $g$ is the determinant of the matrix $g_{\alpha\beta}$.

Let's introduce a new component into our discussion and note that the positive-definite quantity
\begin{equation} \Delta \ell^2\equiv g_{\mu\nu}\Delta a^\mu \Delta a^\nu \label{160}
\end{equation}
has the appearance of a line element or distance between neighboring thermodynamic states \cite{Rupp79}, or more technically, a Riemannian metric, as described in the Appendix. In thermodynamics, distance clearly measures probability: the less the probability of a fluctuation between two states, the further apart they are.

To transform to another pair of thermodynamic coordinates $(x^1,x^2)$, we note that because entropy is a function of state, $\Delta \ell^2$ keeps its value regardless of what coordinates in which it is we expressed. Hence, we can write
\begin{equation} P_G \, dx^1dx^2 = \Big(\frac{V}{2\pi} \Big) \exp \Big(-\frac{V}{2}g_{\mu\nu}\Delta x^\mu \Delta x^\nu \Big)\sqrt{g}\,dx^1dx^2, \label{161}
\end{equation}
provided that we transform the metric elements $g_{\alpha\beta}$ as a second-rank tensor, as in Eq. (\ref{1040}). The tensor transformation property may look novel in thermodynamics, but it is at least implicit in Ref. \cite{Landau}. In general coordinates, the line element is
\begin{equation} \Delta \ell^2\equiv g_{\mu\nu}\Delta x^\mu \Delta x^\nu. \label{162}
\end{equation}

The Gaussian approximation yields the averages
\begin{equation}
\left< \Delta x^{\alpha}\right> =\!\int\! \Delta x^\alpha P_G \, dx^1dx^2 = 0,\label{156}
\end{equation}
\noindent and
\begin{equation}
\left< \Delta x^{\alpha}\Delta x^{\beta}\right>=\!\int\!\Delta x^\alpha \Delta x^\beta P_G \, dx^1dx^2=\frac{g^{\alpha\beta}}{V}, \label{158}
\end{equation}
where $g^{\alpha\beta}$ denotes the components of the inverse of the matrix $g_{\alpha\beta}$.

Particularly simple are coordinates that diagonalize $g_{\alpha\beta}$. For example, in $(T,\rho)$ coordinates \cite{Landau}:
\begin{equation} \Delta \ell^2= \frac{1}{k_B T}\left(\frac{\partial s}{\partial T}\right)_{\rho}\Delta T^2 + \frac{1}{k_B T}\left(\frac{\partial\mu}{\partial\rho}\right)_T\Delta\rho^2. \label{170}
\end{equation}
This line element is readily expressed in terms of the Helmholtz free energy per volume
\begin{equation} f \equiv u-T s.\label{172}
\end{equation}
We have
\begin{equation} s=-\left(\frac{\partial f}{\partial T}\right)_\rho, \label{174}
\end{equation}
\noindent and
\begin{equation}
\mu=\left(\frac{\partial f}{\partial\rho}\right)_T. \label{176}
\end{equation}
Equation (\ref{158}) yields,
\begin{equation}\left< (\Delta\rho)^2\right>=\frac{k_B T}{V}\left(\frac{\partial\rho}{\partial\mu}\right)_T. \label{178}
\end{equation}

For the ideal gas,
\begin{equation} f=\rho k_B T \ln\rho+\rho k_B h(T), \label{182}
\end{equation}
where $h(T)$ is some function of the temperature with negative second derivative to assure a positive heat capacity. Equation (\ref{178}) yields $\left< (\Delta N)^2\right>=\left< N\right>$. Integrating $\Delta T$ out in Eq. (\ref{161}) gives the Gaussian approximation for fluctuations in $\Delta N$. This approximation may be compared with the exact Poisson distribution for $N$. Results for $\left< N\right>=5$ are shown in Fig. 3 \cite{Rupp83a}. For the ideal gas, the Gaussian approximation is effective even at size scales containing just a few particles.

The addition of a second independent fluctuating variable again raises issues of covariance, conservation, and consistency. As in Sec. II, it is easy to construct the correct theory around an appropriate partial differential equation \cite{Rupp95}:
\begin{equation} \frac{\partial P}{\partial t}=-\frac{\partial}{\partial x^{\mu}}\left[K^{\mu}(x)P\right] + \frac{1}{2}\frac{\partial^2}{\partial x^{\mu}\partial x^{\nu}}\left[g^{\mu\nu}(x)P\right]. \label{185}
\end{equation}
In $(u,\rho)$ coordinates, $K^{\alpha}=0$ and $g_{\alpha\beta}$ is given in Eq. (\ref{150}). The coordinate transformations are
\begin{equation} \tilde{P}=P \left| \frac{\partial x}{\partial \tilde{x}} \right|, \label{186}
\end{equation}
where $|\partial x/\partial \tilde{x}|$ is the Jacobian of $x(\tilde{x})$,
\begin{equation}
\tilde{g}_{\mu\nu}=g_{\alpha\beta}\frac{\partial x^{\alpha}}{\partial \tilde{x}^\mu}\frac{\partial x^{\beta}}{\partial \tilde{x}^\nu}, \label{188}
\end{equation}
\noindent and
\begin{equation}
\tilde{K}^{\alpha}= \frac{\partial \tilde{x}^{\alpha}}{\partial x^{\mu}}K^{\mu}+\frac{1}{2}g^{\mu\nu}\frac{\partial^2 \tilde{x}^{\alpha}}{\partial x^{\mu}\partial x^{\nu}}. \label{190}
\end{equation}

For constant $K^{\alpha}$ and constant metric elements $g_{\alpha\beta}$ we can write down the Gaussian expression
\begin{equation} P_G(x,t)=\left(\frac{1}{2\pi t}\right)\sqrt{g} \exp\left[-\frac{1}{2t}\,g_{\mu\nu}(\Delta x^{\mu}-K^{\mu} t)(\Delta x^{\nu}-K^{\nu} t)\right], \label{192}
\end{equation}
which satisfies Eq. (\ref{185}) exactly. Clearly, for small $t$ this Gaussian expression will always be adequate since $K^{\alpha}$ and $g_{\alpha\beta}$ don't get a chance to fluctuate very much. For larger $t$, with its larger fluctuations, the key to getting the Gaussian expression to work is finding a transformation to coordinates with constant $K^{\alpha}$ and $g_{\alpha\beta}$. As in Sec. II, a necessary requirement for this coordinate transformation is that it yield constant $g_{\alpha\beta}$. However, as I will argue in Sec. V, such a coordinate transformation does not generally exist, in contrast to the case with one fluctuating independent variable. There will thus be a lower limit on $V$ where the Gaussian expression must fail regardless of choice of coordinates.

To conclude this section, I point out that Weinhold \cite{Wein} originated thermodynamic metrics in the form of inner products based on the Hessian of the internal energy. The positive-definite nature of this inner product represents the second law of thermodynamics. But Weinhold's geometry lacks a true Riemannian metric structure since it had no underlying physical notion of distance, such as is offered by fluctuations in Eq. (\ref{160}) where the Hessian of the entropy come into play \cite{Rupp79}. An entropy metric was also used by Andresen, Salamon, and Berry \cite{Andres} as a measure of the dissipated availability in finite-time thermodynamics.  Ruppeiner \cite{Rupp83a, Rupp83b} first suggested using a covariant and consistent diffusion type equation for thermodynamic fluctuation theory.  Di\'osi and Luk\'acs \cite{Diosi} completed this theory, adding conservation and drift, and were the first to write the equation in the form in Eq. (\ref{185}).

\section{\label{next}A PHYSICAL LIMIT}

As an application, consider a pure fluid possibly near its critical point. The attractive part of the force between particles causes particles to be closer to each other on the average than they would be in a purely random situation. A clustered pair will attract a third particle, and so on, resulting in a droplet of particles of some characteristic size, the correlation length $\xi$ \cite{Pathria}. This length is zero for the ideal gas and approaches infinity at the critical point. Unless we are in a regime where the fluid coexists in two phases (assumed not to be the case), these droplets continually disassociate and reform.

This picture is an oversimplification. The density of the fluid is a function of position $\rho=\rho(\vec{r})$, which has contour surfaces of constant density. Consider a surface of constant $\rho_0$. According to Widom \cite{Widom}, this surface is enormously complex and sponge-like. Any straight line makes many intersections with it, as shown in Fig. 4 (reproduced from Widom \cite{Widom}). The mean distance between these intersections is a measure of $\xi$. Because these intersections occur independent of the direction of the line, it is convenient to think of the regions $\rho<\rho_0$ and $\rho>\rho_0$ as each consisting of many small, independent volume elements (''droplets'') of linear dimension $\xi$ (see Fig. 4) though of course these volume elements are in reality connected in one sponge-like mass.

The correlation length $\xi$ is commonly assumed to require statistical mechanics for its proper definition and calculation. The key argument of this paper is that $\xi$ can be deduced from thermodynamics. Notice in Fig. 4 that if $V\gg\xi^3$, then system $A$ would see itself mostly surrounded by a local environment with density near to $\rho_0$ at any time. Being able to regard the reservoir $A_r$ of $A$ as a uniform system with density $\rho_0$ is an essential requirement in the Gaussian expression, and it is thus justified for $V\gg\xi^3$. If, however, $V\ll\xi^3$, $A$ would at any time most likely be in a region between the curves of constant density $\rho_0$ in Fig. 4. The local surroundings now seen by $A$ at any time has a thermodynamic state different from that of the overall system $A_0$. The density fluctuations would be typically bimodal, as in Fig. 5.

The length scale $\xi$ thus marks the intrinsic breakdown of the Gaussian expression. In Sec. \ref{next2}, I relate this breakdown to the mathematical structure we have been developing.

\section{\label{next2}THERMODYNAMIC CURVATURE}

For $V$ sufficiently large, the physical limit discussed in Sec. \ref{next} does not constrain us, and in $(u,\rho)$ coordinates the Gaussian expression should work just fine. For small fluctuations, the matrix elements $g_{\alpha\beta}$ will be nearly constant over the full range of fluctuations.

But even in this apparently safe volume regime we can always invalidate the Gaussian expression by transforming to ``bad'' coordinates in which $g_{\alpha\beta}$ varies sharply for even small fluctuations. However, this mathematical game has no relation to anything physically fundamental. We should simply avoid such coordinates. It is far more interesting to transform to ``good'' coordinates with the Gaussian expression extended to small $V$. As we have seen, a necessary property of good coordinates is slowly varying $g_{\alpha\beta}$ as $A$ fluctuates. Better yet, might we find ``great'' coordinates with the Gaussian expression extended to near atomic volumes? A reason to hope for ``great" are the ideal gas results shown in Fig. 3, where we lucked into such coordinates. But the ideal gas, with $\xi=0$, does not suffer from the physical breakdown of Sec. \ref{next}, and thus it is no surprise that the Gaussian expression works to such small volumes. More generally, $\xi$ marks the physical onset of non-Gaussian fluctuations in all coordinates. I will now argue that this volume can be deduced from thermodynamics.

The key is the Riemannian curvature scalar $R$ of the metric $g_{\alpha\beta}$. Imagine $V$ large enough for the Gaussian expression Eq. (\ref{192}) to be valid in some choice of coordinates. For the Gaussian expression to be valid, we require constant $g_{\alpha\beta}$ over typical fluctuations. With constant $g_{\alpha\beta}$, simple linear algebra allows a transformation to a locally flat Cartesian line element, as in Eq. (\ref{1070}). Clearly, a valid Gaussian expression requires the system to be in the locally flat geometric regime.

This relation is elementary. Much more interesting are larger fluctuations. Define the squared distance
\begin{equation}
r^2\equiv g_{\mu\nu}\Delta x^\mu \Delta x^\nu \label{198}
\end{equation}
to be the size of a typical fluctuation of the state of $A$ about that of $A_r$. The argument leading to Eq. (\ref{1080}) leads us to the observation that
\begin{equation}
r^2\sim \frac{12}{|R|}, \label{199}
\end{equation}
beyond which we are no longer in the geometric locally flat regime. Beyond this approximate value of $r$, no local coordinate system with constant $g_{\alpha\beta}$ exists, and the Gaussian expression fails in all coordinate systems.

The argument of the exponential in Eq. (\ref{161}) is roughly $-1$ for typical fluctuations, corresponding to
\begin{equation} r^2 \sim\frac{2}{V}. \label{200}
\end{equation}
If we combine Eq. (\ref{200}) with Eq. (\ref{199}), we see that the Gaussian expression must fail in all coordinate systems if
\begin{equation} V < \sim \frac{|R|}{6}.\label{220}
\end{equation}

This breakdown is intrinsic to the geometry, and has nothing to do with picking bad coordinates. Nor may it be solved by any good coordinates. It calls for some physical interpretation. I relate it to the physical breakdown of the Gaussian expression in Sec. IV and interpret $|R|$ as being proportional to the correlation volume
\begin{equation}
|R|\sim \xi^d,\label{250}
\end{equation}
where $d$ is the dimensionality of space.

\section{CALCULATION OF $R$}

Given some statistical mechanical model, we may test the proposition that $|R|\sim \xi^d$ by calculating both $R$ and $\xi$. We first evaluate $R$ for the ideal gas and then follow with a brief review of numerous other models for which $R$ has been evaluated. I also make some observations about the sign of $R$.

Although $R$ for the ideal classical gas requires almost no calculation, this calculation nevertheless makes a good exercise. We already know $|R|$ must be small because if particles move randomly with respect to each other, there is no tendency for them to cluster. Equations (\ref{170})-- (\ref{176}) and (\ref{182}) yield the line element in coordinates $\{x^1,x^2\}=\{T,\rho\}$:
\begin{equation} \Delta \ell^2=-\frac{\rho h''(T)}{T}\Delta T^2 + \frac{1}{\rho}\Delta\rho^2, \label{252}
\end{equation}
from which we can read off the metric elements
\begin{equation} \left \{g_{11},g_{22} \right\}= \left\{-\frac{\rho h''(T)}{T},\frac{1}{\rho} \right\}. \label{253}
\end{equation}
We substitute these metric elements into Eq. (\ref{1140}) for $R$ and find the exact result
\begin{equation}
R=0 \label{254}
\end{equation}
for the ideal classical gas \cite{Rupp79}.

This result came as a surprise, and pointed to some mysterious connection between curvature and interactions. But the argument given in Sec. V clears up the mystery. Also initially surprising but now understood is the proportionality $|R|\sim \xi^d$ in Eq. (\ref{250}). It has been confirmed for several statistical mechanics models \cite{Rupp79, Rupp81, Rupp90B, Brody95, Dol97, Dol02, Jan02, Jan03, Brody03, John}, particularly near critical points, where $\xi^d$ is large enough to encompass a volume containing many particles. I add that $R$ is typically negative in my sign convention (see Appendix) in the standard critical point models, for example, van der Waals for which the attractive interaction dominates the long-range behavior.

Another key example is an ideal gas of bosons. This system has no interparticle interactions, but quantum statistics leads particles to bunch closer than in a classical ideal gas. This clustering should be measured by $R$. The ideal Bose gas suffers a Bose-Einstein condensation where a macroscopic number of particles pool in the ground state at low temperature. It was found that $R$ is negative and diverges to negative infinity at absolute zero \cite{Mrug90}.

Another example is an ideal gas of fermions. $R$ was found to be positive and to diverge to positive infinity at absolute zero \cite{Mrug90,Oshima}. Janyszek and Mruga{\l}a \cite{Mrug90} have emphasized the difference in the sign of $R$ between the Bose and Fermi gases. An explicit connection with the correlation length for ideal quantum gasses remains to be established.

Table \ref{tab1} shows the sign and divergence of $R$ for several systems \cite{Rupp08}. $R$ is negative where attractive interactions dominate, and positive $R$ where repulsive interactions dominate \cite{footnote3}. Table \ref{tab1} includes several weakly interacting systems with $R$ on the order of the volume of an interparticle spacing or less. I interpret such values of $R$ as physically equivalent to zero, because the meaning of $\xi^d$ of this size is lost in the noise as thermodynamics breaks down as individual particles or spins become visible. The one-dimensional antiferromagnetic Ising model \cite{Rupp81,Mrug} may be misplaced in this category, since the ordering field in antiferromagnetic systems is a staggered field, and not the constant field used for the calculations in Table \ref{tab1}.

There are four systems in Table \ref{tab1} for which $R$ is both positive and negative. Not all these have been analyzed in terms of attractive and repulsive interactions. The sign of $R$ for the one-dimensional $q$-state Potts model \cite{Dol02, John} is related to $q$. For $q>2$ and nonzero magnetic field, there are significant regions of negative $R$ at low temperatures. The Takahashi gas \cite{Rupp90B} has negative $R$ in the gas-like phase where attractive interactions dominate, and small $|R|$ in the liquid-like phase where interactions are effectively short-range. Going from one phase to the other by changing the density at constant temperature gives a pseudophase transition accompanied by a sharp positive spike in $R$. An abrupt change in the sign of $R$ is also present in the one-dimensional Ising ferromagnet of finite $N$ spins \cite{Brody03}. $R$ is negative for large $N$, and sharply increases to large positive values as $N$ is decreased. Both positive and negative signs for $R$ are present in the noninteracting gas with Gentile's statistics \cite{Oshima}.  Such a gas allows a maximum number of particles $p$ into any quantum state.  In the classical limit $|R|$ is always small, $R$ is negative for $p\ge 2$, and $R$ is positive for the Fermi gas ($p=1$).  In the degenerate limit  $T\rightarrow 0$, $R\rightarrow +\infty$ for all finite values of $p$.  Oshima et al. \cite{Oshima} also corrected some earlier Fermi gas results \cite{Mrug90}. Also in the category of having $R$ with both signs is a gas of anyons \cite{Mirza}, particles in two dimensions with fractional spin. This model has a parameter whose variation allows us to change it continuously from a Bose gas to a Fermi gas; the sign of $R$ changes correspondingly from negative to positive.

\section{CONCLUSIONS}

Can thermodynamic curvature tell us anything we could not obtain otherwise? If we have a microscopic model, statistical mechanics offers powerful methods for calculating its thermodynamic properties. It also usually allows us to calculate microscopic properties, such as the correlation length. Calculating $R$ might not teach us much about such a model.

If the system's underlying microscopic structure is unknown, but we know its thermodynamic properties, calculating $R$ gives a bigger payoff. Such a case occurs for black holes which has a well developed thermodynamics \cite{Beck80}. Little is known about its possible microscopic foundations. There have been several calculations of the thermodynamic curvature $R$ for black holes \cite{Rupp08,Aman}. It has been found \cite{Rupp08} that the sign of $R$ is almost always positive for the spinning, charged Kerr-Newman black hole, suggesting that an appropriate microscopic model might consist of some type of repulsive fermions rather than one of the standard critical point models listed in Table I.

If $R$ does augment thermodynamics in some fundamental way, then we must investigate why and to what extent. For example, we have interpreted just $|R|$. Beyond the observations summarized in Table \ref{tab1}, it is not clear what the sign of $R$ signifies and why.

$|R|$ relates to hyperscaling in critical phenomena \cite{Pathria, Widom}. Hyperscaling relates an appropriate free energy $\phi$ to the correlation length by $\phi\sim\xi^{-d}$. Combining this with $|R|\sim \xi^d$ yields $|R|\sim\phi^{-1}$ \cite{Rupp91}. Because $R$ may be expressed in terms of $\phi$ and its derivatives, we have a partial differential equation for $\phi$. Its solution, in conjunction with boundary conditions, yields $\phi$ using no statistical mechanics at all. Tests have been few, so any review of results here would be premature.

\section{Appendix: Riemannian geometry}

Consider a two-dimensional surface embedded in a three-dimensional Euclidean space with Cartesian coordinates $(x,y,z)$ via some function $z=z(x,y)$. For example, a two-sphere centered at the origin with radius $a$ gives $z=\pm\sqrt{a^2-x^2-y^2}$, as in Fig. 6.

The two-dimensional surface, with some choice of coordinates $(x^1,x^2)$, inherits its distance from the surrounding Euclidean space. Its line element is \cite{Arfken}
\begin{equation} \Delta \ell^2\equiv g_{\mu\nu}\Delta x^\mu \Delta x^\nu, \label{1010}
\end{equation}
where $\mu,\nu=1,2,$ and the metric elements $g_{\alpha\beta}$ constitute a positive-definite matrix. For example, the two-sphere in spherical coordinates $(\theta,\phi)$ has
\begin{equation} \Delta \ell^2=a^2 \Delta\theta^2+ a^2\,\sin^2 \theta\,\Delta\phi^2. \label{1000} \end{equation}

The distance between pairs of neighboring points does not depend on the choice of coordinate system, and hence $\Delta \ell^2$ transforms as a scalar. Consider some other pair of coordinates $(\tilde{x}^1,\tilde{x}^2)$ in which the metric elements are $\tilde{g}_{\alpha\beta}$. Because
\begin{equation} \Delta x^\alpha=\frac{\partial x^{\alpha}}{\partial \tilde{x}^\mu} \Delta \tilde{x}^\mu,\label{1020}
\end{equation}
we have
\begin{equation}\Delta \ell^2=g_{\alpha\beta}\frac{\partial x^{\alpha}}{\partial \tilde{x}^\mu}\frac{\partial x^{\beta}}{\partial \tilde{x}^\nu}\Delta \tilde{x}^\mu \Delta \tilde{x}^\nu\equiv \tilde{g}_{\mu\nu} \Delta \tilde{x}^\mu \Delta \tilde{x}^\nu.\label{1030} \end{equation}
The transformed metric elements are thus
\begin{equation} \tilde{g}_{\mu\nu}=g_{\alpha\beta}\frac{\partial x^{\alpha}}{\partial \tilde{x}^\mu}\frac{\partial x^{\beta}}{\partial \tilde{x}^\nu}. \label{1040}
\end{equation}
Equation (\ref{1040}) is the transformation rule for a second-rank tensor \cite{Arfken, Laug}. The two-dimensional surface also has area element $\sqrt{g}dx^1dx^2$ which transforms to $\sqrt{\tilde{g}}d\tilde{x}^1d\tilde{x}^2$, with $g$ and $\tilde{g}$ the determinants of $g_{\alpha\beta}$ and $\tilde{g}_{\alpha\beta}$, respectively \cite{Weinberg}.

We measure distance along any curve on the two-dimensional surface by integrating the line element along it:
\begin{equation} \int \sqrt{g_{\mu\nu} dx^\mu dx^\nu}.\label{1050}
\end{equation}
Introduce curves on the two-dimensional surface minimizing the distance between pairs of points. These curves are called geodesics. Imagine a simple exercise starting from some central point anywhere on the two-dimensional surface and drawing along geodesics a fixed distance $r$ (not too large) in all directions. The resulting endpoints define a circle; see Fig. 6 for the circle on the two-sphere.

Locally, any two-dimensional surface looks flat, and a very small circle has circumference $C \approx 2\pi r$. Draw out further and deviations from locally Euclidean geometry appear. For example, for the circle on the two-sphere in Fig. 6, $C$ is less than $2\pi r$ because the circle lies entirely on one side of a plane tangent to the two-sphere at the center of the circle. A challenging exercise in non-Euclidean geometry shows that \cite{Laug}
\begin{equation}
C = 2\pi r + \frac{\pi}{6} R r^3 + O(r^4),\label{1060}
\end{equation}
where the Riemannian curvature scalar $R$ gives the size of the leading correction to Euclidean geometry.

The value of $R$ depends on where we are on the surface. (Historically, the Gaussian curvature $K=-R/2$ came first. $K$ is unambiguously defined and is positive for the two-sphere. The sign of $R$ is subject to convention. I use Weinberg's sign convention \cite{Weinberg}, with $R$ negative for the two-sphere.) Because $R$ is based on lengths of curves, it is independent of the choice of coordinate system, and is a purely geometrical quantity. For a plane $R=0$. For the two-sphere in Fig. 6 a simple calculation shows $R=-2/a^2$. For a diagonal metric we have \cite{Laug}
\begin{equation}
R=\frac{1}{\sqrt{g}}\left[\frac{\partial}{\partial x^1} \left(\frac{1}{\sqrt{g}}\frac{\partial g_{22}}{\partial x^1}\right)+\frac{\partial}{\partial x^2}\left(\frac{1}{\sqrt{g}}\frac{\partial g_{11}}{\partial x^2}\right)\right].\label{1140}
\end{equation}

Surfaces of constant positive $R$ also exist, pseudospheres as in Fig. 7 \cite{Hsiung}. Here, a small circle drawn out from any point lies on both sides of a plane tangent to the two-sphere at the center of the circle, and $C$ is larger than that of a circle with the same radius drawn on a plane.

For simplicity, I have emphasized drawing small circles. Other questions may be asked regarding deviations from Euclidean geometry. For example, how much does the sum of the interior angles of a triangle with geodesic sides deviate from $\pi$? $R$ answers all such questions.

In a small neighborhood of a non-singular point, the geometry is always flat. Hence, we may find a pair of locally Cartesian coordinates $(X,Y)$ on the two-dimensional surface with
\begin{equation}
\Delta \ell^2= \Delta X^2+ \Delta Y^2. \label{1070}
\end{equation}
How large may $r$ become before ``flat'' becomes inadequate? For a rough estimate, set the correction term in Eq. (\ref{1060}) equal to the leading term, yielding
\begin{equation} r^2\sim \frac{12}{|R|}.\label{1080} \end{equation}
For $r$ beyond this limit, the geometry is intrinsically non-Euclidean, and may not be represented with locally Cartesian coordinates because they will always give $2\pi r$ for the circle circumference. Neither is it possible to find any local coordinate system in which we can treat $g_{\alpha\beta}$ as effectively constant. If such a coordinate system existed, we could transform these coordinates to locally Cartesian coordinates, which we know cannot represent the curvature of the surface.

The discussion has been of the surface geometry originated by Gauss \cite{Gauss} which starts with an embedding function $z(x,y)$, and a surface metric induced by the embedding in the three-dimensional Euclidean space. The approach in thermodynamics is the converse. We start with a surface metric, and seek a corresponding embedding function. But this converse problem usually has no solution, and an embedding function rarely exists \cite{Andres88}. Therefore, Gauss's surface geometry is in most cases useless in metric thermodynamics.

To solve this problem, we need to go one step further to Riemann's geometry. Riemann kept as basic the notion of a smooth, two-dimensional surface of points with coordinates $(x^1,x^2)$. (More technically, a two-dimensional manifold. Generalization to higher dimensions is possible.) He dropped the embedding in a three-dimensional Euclidean space, and with it the function $z(x,y)$, but kept the line element Eq. (\ref{1010}), lengths of curves Eq. (\ref{1050}), and deviations from Euclidean geometry characterized by measurements within the two-dimensional surface, such as those leading to Eq. (\ref{1060}). Riemann also kept equations independent of embedding, such as Eq. (\ref{1140}) for $R$. Thermodynamic metric geometry requires a switch to Riemann's picture. However, for visualizations and calculations we are usually safe using the ideas and many relations from Gauss' geometry.

I thank Harvey Gould for many suggestions which improved the narrative flow.

\newpage

\section*{Table}

\begin{table}[h!]
\centering
\begin{tabular}{l|c|c|c|l}
\hline
System & $n$ & $d$ & Sign of $R$ & Divergence \\
\hline
3D Bose gas \cite{Mrug90} & $2$ & $3$ & $-$ & $T\rightarrow 0$ \\
1D Ising ferromagnet \cite{Rupp81,Mrug} & $2$ & $1$ & $-$ & $T\rightarrow 0$ \\
Critical region \cite{Rupp95,Rupp79,Brody95} & $2$ & $\cdots$ & $-$ & critical point \\
Mean-field theory \cite{Mrug} & $2$ & $\cdots$ & $-$ & critical point \\
van der Waals \cite{Rupp95,Brody95,Brody09} & $2$ & $3$ & $-$ & critical point \\
Ising on Bethe lattice \cite{Dol97} & $2$ & $\cdots$ & $-$ & critical point \\
Ising on 2D random graph \cite{Jan02,John} & $2$ & $2$ & $-$ & critical point \\
Spherical model \cite{Jan03,John} & $2$ & $\cdots$ & $-$ & critical point \\
Self-gravitating gas \cite{Rupp96} & $2$ & $3$ & $-$ & unclear \\
1D Ising antiferromagnet \cite{Rupp81,Mrug} & $2$ & $1$ & $-$ & $|R|$ small \\
Tonks gas \cite{Rupp90B} & $2$ & $1$ & $-$ & $|R|$ small \\
Ideal gas \cite{Rupp79} & $2$ & $3$ & $0$ & $|R|$ small \\
Ideal paramagnet \cite{Rupp81,Mrug} & $2$ & $\cdots$ & $0$ & $|R|$ small \\
Multicomponent ideal gas \cite{Rupp90} & $>2$ & $3$ & $+$ & $|R|$ small \\
1D Potts model \cite{Dol02,John} & $2$ & $1$ & $\pm$ & $T\rightarrow 0$ \\
Takahashi gas \cite{Rupp90B} & $2$ & $1$ & $\pm$ & $T\rightarrow 0$ \\
Finite 1D Ising ferromagnet \cite{Brody03} & $2$ & $1$ & $\pm$ & $T\rightarrow 0$ \\
Anyon gas \cite{Mirza} & $2$ & $2$ & $\pm$ & $T\rightarrow 0$ \\
2D Fermi gas \cite{Rupp08} & $2$ & $2$ & $+$ & $T\rightarrow 0$ \\
3D Fermi gas \cite{Mrug90,Oshima} & $2$ & $3$ & $+$ & $T\rightarrow 0$ \\
Gentile's statistics (finite $p$)\cite{Oshima} &$2$&$3$&$+$ & $T\rightarrow 0$ \\
3D Fermi paramagnet \cite{Kav} & $3$ & $3$ & $+$ & $T\rightarrow 0$ \\
\hline
\end{tabular}
\caption{\label{tab1}The thermodynamic curvature for several thermodynamic systems. The number of independent thermodynamic parameters $n$, spatial dimension $d$, sign of $R$, and possible divergences is given. For some systems $d$ is not determined, which is denoted by $\cdots$. The sign of $R$ is consistent with the sign convention of Weinberg \cite{Weinberg}. Small $|R|$ means that the value of $|R|$ is on the order of the volume of an interparticle spacing or less.}
\end{table}

\clearpage\section*{Figure Captions}

\begin{figure}[h!]
\caption{An imaginary partition inside an infinite, closed universe $A_0$. The partition encloses a system $A$, with constant volume V. $A_r$ is the reservoir of $A$, which is everything in $A_0$ outside the partition.}
\end{figure}

\begin{figure}[h!]
\caption{Two concentric, imaginary partitions enclosing systems $A_1$ and $A$, respectively ($A$ is part of $A_1$). $A_r$ is now everything in $A_0$ outside of $A_1$. $A_1$ samples the state of $A_0$, but $A$ samples only the present state of $A_1$.}
\end{figure}

\begin{figure}[h!]
\caption{The ideal gas probability $P(N)$ of finding $N$ particles in $A$ for $V$ such that $\left< N\right>=5$. $P(N)$ is calculated with the Gaussian approximation (solid curve) and the exact Poison distribution
(points). The Gaussian approximation is effective even for $V$ close to atomic sizes.}
\end{figure}

\begin{figure}[h!]
\caption{A surface on which the mean local density $\rho$ equals the overall mean density $\rho_0$. Also shown is an arbitrary line intersecting the surface at the dotted points, and a ''droplet'' of linear dimension $\xi$ equal to the mean distance between those intersections.}
\end{figure}

\begin{figure}[h!]
\caption{For large $V$ the density fluctuations are given by a Gaussian probability $P(\rho)$ centered on $\rho_0$. If $V$ is less than $\xi$, fluctuations are bimodal, because $A$ is most probably either inside a dense cluster or in a depleted in-between region.}
\end{figure}

\begin{figure}[h!]
\caption{A two-sphere with radius $a$. A circle centered at the north pole is drawn on it. Its radius $r$ measured along a geodesic on the two-sphere is greater than its radius measured in the flat three-dimensional embedding space, so its Riemannian curvature scalar $R$ is negative.}
\end{figure}

\begin{figure}[h!]
\caption{The pseudosphere defined by the parametric functions $x=a \sin u \cos v$, $y=a \sin u \sin v$, and $z=a[\cos u + \ln\tan (u/2)]$, has constant curvature $R=2/a^2$.}
\end{figure}

\includegraphics[width=6in]{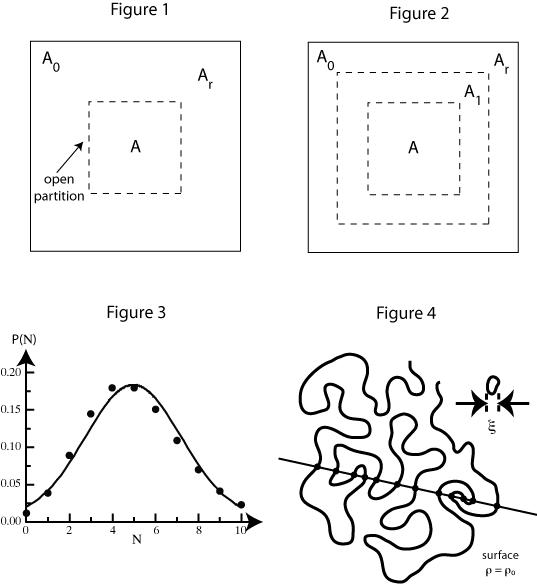}
\includegraphics[width=6in]{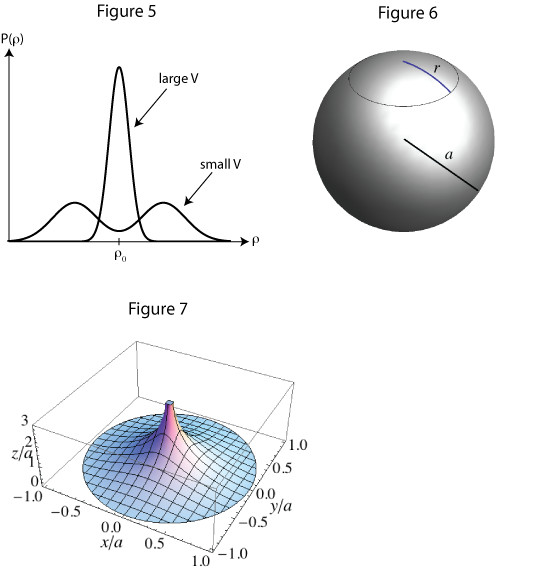}

\end{document}